\documentstyle[11pt, aaspp4, psfig]{article}

\begin{document} 
\title{Oscillation Waveforms and Amplitudes from Hot Spots on Neutron Stars}

\author{Nevin Weinberg}
\affil{Department of Astronomy and Astrophysics, University of Chicago\\
       5640 South Ellis Avenue, Chicago, IL 60637\\
       nweinber@midway.uchicago.edu}

\author{M. Coleman Miller}
\affil{Department of Astronomy, University of Maryland\\
       College Park, MD  20742-2421\\
       miller@astro.umd.edu}

\author{Donald Q. Lamb}
\affil{Department of Astronomy and Astrophysics and Enrico Fermi Institute\\
       University of Chicago, 5640 South Ellis Avenue, Chicago, IL 60637\\
       lamb@pion.uchicago.edu}

\begin{abstract} 
The discovery of high-amplitude brightness oscillations during
type I X-ray bursts from six low-mass X-ray binaries has provided a powerful
new tool to study the properties of matter at supranuclear densities, the
effects of  strong gravity, and the propagation of thermonuclear burning. There
is substantial evidence that these brightness oscillations are produced by
spin  modulation of one or two localized hot spots confined to the stellar
surface. It is therefore important to calculate the expected light curves
produced  by such hot spots under various physical assumptions, so that
comparison   with the observed light curves may most sensitively yield
information about the underlying physical quantities. In this paper we make
general relativistic calculations of the light curves and oscillation
amplitudes produced by a  rotating neutron star with one or two hot spots as a
function of spot size,  stellar compactness, rotational velocity at the stellar
surface, spot location, orientation of the line of sight of the observer, and
the angular dependence of the surface specific intensity. For the case of two
emitting spots we  also investigate the effects of having spot separations less
than $180^\circ$ and the effects of having asymmetries in the brightness of the
two spots.  We find that stellar rotation and beaming of the emission tend to
increase the observed  oscillation amplitudes whereas greater compactness and
larger spot size tend to decrease them. We also show that when two emitting
spots are either non-antipodal or asymmetric in brightness,  significant power
at the first harmonic is generated. By applying these results to 
4U~1636--536, whose two emitting spots produce power at the first harmonic, we 
place strong constraints on the neutron star's magnetic field
geometry.  We also show that the data on the phase lags between photons
of different energies in the
persistent pulsations in SAX~J1808--58 can be fit well with a model in which
the observed hard leads are due to Doppler beaming.
\end{abstract}

\keywords{stars: neutron --- equation of state --- gravitation --- 
relativity --- X-rays: bursts}

\section{INTRODUCTION}

The study of neutron stars is attractive in part because of the fundamental
issues of physics that can be addressed.  These include the behavior of
spacetime in strong gravity, the equation of state of matter at supranuclear
densities, and the propagation of thermonuclear burning in degenerate
matter, an issue which has relevance to many astrophysical events including
classical novae and Type~Ia supernovae.

The discovery with the \textsl{Rossi X-ray Timing Explorer} (\textsl{RXTE}) of
highly coherent brightness oscillations during type I (thermonuclear) X-ray 
bursts from six low mass X-ray binaries (LMXB) (for reviews see, e.g., 
Strohmayer, Zhang \& Swank 1997b; Smith, Morgan \& Bradt 1997; Zhang et al.\
1996; and Strohmayer et al.\ 1997c) has provided a potentially sensitive new
tool to understand these fundamental issues.  The burst oscillations are thought
to be produced by spin modulation of one or two localized thermonuclear hot
spots that are brighter than the surrounding surface.  The existence of
the oscillations, as well as some of the reported behavior of their
amplitudes (see, e.g., Strohmayer et al.\ 1997b) seems to confirm the
pre-existing theoretical expectation that X-ray bursts on neutron stars
are caused by ignition at a point followed by thermonuclear propagation
around the surface (e.g., Fryxell \& Woosley 1982; Nozakura et al. 1984; 
Bildsten 1995).  The observed 
waveforms of these oscillations, and their dependence on time and photon
energy, can in principle be used to constrain
the mass and radius of the star and the velocity and type of
thermonuclear propagation.  Such information can only be extracted by
detailed comparison of theoretical waveforms with the data.

Here we conduct the most complete existing survey of the properties of the
light curves and resultant oscillation amplitudes for one or two expanding hot
spots. We calculate light curves and oscillation amplitudes as a function of
stellar compactness, rotational velocity at the stellar surface, spot size
and location, orientation of the line of sight, angular dependence of the
specific intensity, and spot asymmetries.
Our calculations follow a procedure similar to that of
Pechenick, Ftaclas, \& Cohen (1983), Strohmayer (1992), and Miller \& Lamb
(1998), but
our survey is more comprehensive than these previous treatments in that
we fully investigate the effects of an expanding spot size on the light curves
and oscillation amplitudes, while also exploring the effects of gravity,
stellar rotation, viewing geometries, and anisotropic emission.  In addition,
we present the first calculations of the effects of having two non-antipodal 
spots as well as the  effects of asymmetries in spot brightness. 

In \S~2 we describe our assumptions and the calculational method. In \S~3 we
present our results. We show that for small spot sizes the
oscillation amplitude has only a weak dependence on spot size, but that as the
spot grows the dependence becomes very strong. We also show that stellar
rotation, beaming functions and spot asymmetries all tend  to increase the
observed oscillation amplitudes whereas greater compactness and larger spot
sizes tend to decrease the amplitudes. In \S~4 we exhibit applications of
these results to data on the amplitudes of two harmonics in 4U~1636--536
and on the phase lags versus energy for SAX~J1808--3658.
We discuss our results and present our conclusions in \S~5.

\section{CALCULATIONAL METHOD}

We make the following assumptions in our calculations: 
\begin{enumerate} 
\item \textsl{The observed radiation comes from one or two emitting spots 
on the surface.}
The sources with strong bursts tend to have persistent accretion rates
a factor of $\sim$10--100 less than the Eddington critical luminosity $L_E$
at which the radial radiation force balances gravity,
whereas the peak luminosity of the bursts is typically close to $L_E$.
The flux from the burning regions therefore greatly exceeds the flux from
the quiescent regions, so for much of the burst this is a plausible
approximation. 
\item \textsl{The radiation is homogeneous and emitted isotropically unless 
noted otherwise.} 
This assumption is made for simplicity, as presently there  is no
physical evidence that suggests whether or not the photon emission  from the
hot spots is isotropic and homogeneous.  
\item \textsl{If there are two spots, they
are identical and both grow at the same velocity unless noted otherwise.}
This assumption is also made for simplicity.
Although the geometry of the two magnetic poles is unlikely to be identical,
not enough is known about their structure to realistically model non-identical
spots.  
\item \textsl{The exterior spacetime of the neutron star is the
Schwarzschild spacetime.} 
We neglect the effect of frame dragging due to stellar
rotation because it only generates small second order effects for the rotation
rates of interest (see Lamb \& Miller 1995 and Miller \& Lamb 1996).
\end{enumerate}

\begin{figure*} \hbox{\hskip 1.8truein
\psfig{file=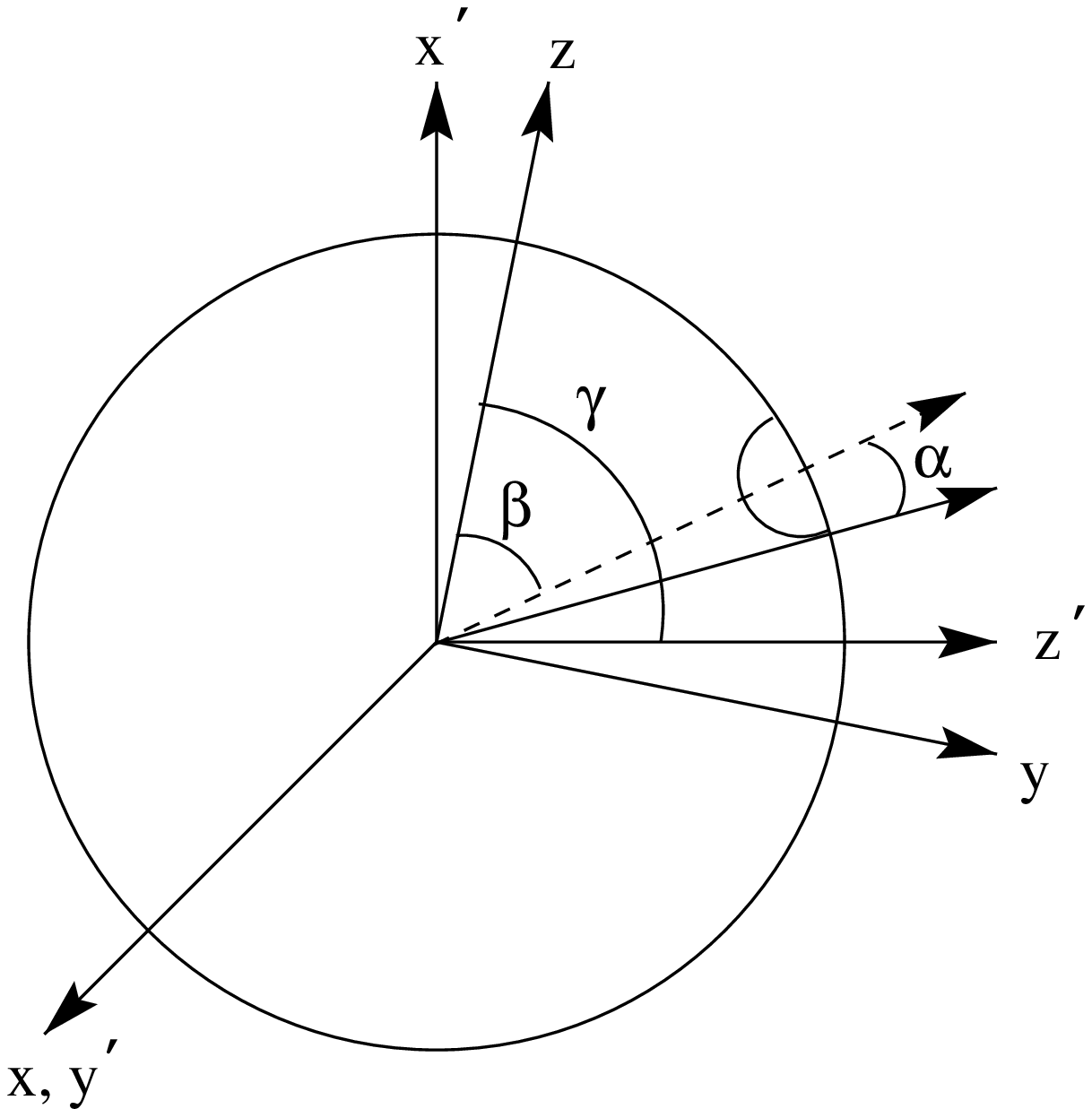,height=3.0truein,width=3.0truein}}
\caption[]{\label{coords}
Coordinate systems used in this paper.  The unprimed axes
correspond to a frame fixed with respect to the star, in which the
$z$-axis is the stellar rotation axis.  The hot spot is fixed on 
the star and rotates with it; the center of the hot spot is an angle
$\beta$ from the rotation axis, and the angular radius of the hot spot
is $\alpha$.  The primed axes are for a coordinate system at rest
with respect to the observer at infinity, who is in the direction corresponding
to the positive $z^\prime$ axis, at an angle $\gamma$ from the rotation
axis.  This figure is based on a similar figure in Strohmayer (1992).} 
\end{figure*}

We compute the waveform of the oscillation as seen at infinity using
the procedure of Pechenick et
al.\ (1983). Figure~\ref{coords} 
shows our coordinate system and angle definitions.
The photons emitted from the star travel along null geodesics which, for a
Schwarzschild geometry, satisfy the equation (Misner, Thorne, and Wheeler 1973,
p. 673)
\begin{equation} \left( \frac{1}{r^2} \frac{dr}{d\phi} \right)^2 + 
\left( \frac{1-2M/r}{r^2} \right)  =  \frac{1}{b^2} 
\end{equation} 
where $r$
and $\phi$ are spherical coordinates, $M$ is the gravitational
mass of the star,  and $b$ is
the impact parameter of the photon orbit. In both the above equation and
throughout, we use geometrized units in which $G=c\equiv 1$.  If the photon
is initially at a global azimuthal angle $\phi=0$, then the global
azimuthal angle at infinity follows from
equation (1) and is (Pechenick et al. 1983, eq. [2.12]) 
\begin{equation}  
\phi_{\rm obs} =
\int_{0}^{M/R} \left[ u_b^2 -  \left(1-2u\right)u^2 \right]^{-1/2} \,du
\end{equation} 
where $u_b = M/b$.  Note that not
all of this angle is due to light deflection: for example, a photon
emitted tangent to the radial vector in flat spacetime will have an
angle $\phi_{\rm obs}=\pi/2$ at infinity.  The maximum angle
occurs when $b = b_{max} = R(1-2M/R)^{-1/2}$ and is given by (Pechenick et al.
1983, eq. [2.13]) 
\begin{equation}\label{eq:maxphi} \phi_{max} = \int_{0}^{M/R}
\left[ \left(1-\frac{2M}{R}\right) \left(\frac{M}{R}\right)^2 -
\left(1-2u\right)u^2 \right]^{-1/2} \,du 
\end{equation}   	 
The observer at infinity cannot see the spot if the observer's azimuthal
angle exceeds $\phi_{\rm max}$.

For each phase of rotation we compute the projected area of many small elements
of a given finite size spot. We then build up the light curve of the entire
spot by superposing the light curve of all the small elements.
We chose a grid resolution such
that the effect of having a finite number of small elements produces a 
fractional error $<10^{-4}$ in the computed oscillation amplitudes.
For isotropic emission the intensity of radiation at a given
rotational phase as seen by an observer at infinity is directly proportional to
the projected area of the spot. To investigate the effect of anisotropic
emission we include a flux distribution function in the intensity, $f(\delta)$,
where $\delta$ is the angle between the surface normal and the photon  
propagation
direction. The intensity is then proportional to the product of the projected
area of the spot (which is proportional to $\cos\delta$) 
and $f(\delta)$. We consider two types of  anisotropic
emission: cosine (``pencil") beaming, in which  $f(\delta) = \cos \delta$, and
sine (``fan") beaming, in which $f(\delta)  = \sin \delta$.  

The intensity distribution of an
emitting spot is aberrated by the rotation of the star, 
and the photon frequency is Doppler shifted by the
factor $1/[\gamma(1-v \cos \zeta)]$. Here $v$ is the velocity at the stellar
equator, $\gamma = (1 - v^2)^{-1/2}$,  and $\zeta$ is the angle between the
direction of photon propagation and the local direction of rotation.  The
inferred spin frequencies of these neutron stars are $\sim$300~Hz, implying
surface velocities $v\sim 0.1$c for stellar radii $R\sim 10$~km.

After computing the oscillation waveform using the above
approach, we Fourier-analyze the resulting light curve to determine the
oscillation amplitudes and phases as a function of photon energy
at different harmonics.

\section{RESULTS}

As discussed in the introduction, the basic quantities of interest include
the mass and radius of the neutron stars in bursting sources and the nature 
and speed of thermonuclear propagation on the stellar surface.  We therefore
need to relate these fundamental quantities to the observables,
such as the oscillation waveform as a function of time and photon energy.
We do this by computing theoretical waveforms using different assumptions
about the compactness of the star, the angular size of the burning region,
the angular location of the observer and magnetic pole relative to the
stellar rotation axis, the surface rotation velocity of the star, and
the angular distribution of the specific intensity at the surface.  In
this section we consider each of these effects separately, to isolate
the effect they have on the waveform and facilitate interpretation of the
data.  Here we always quote the fractional rms amplitude of brightness
oscillations.  We also quote only bolometric amplitudes in this section;
as shown by Miller \& Lamb (1998), oscillations in the energy spectrum of
the source may yield substantially higher amplitudes in the countrate
spectrum measured by bandpass-limited instruments such as {\it RXTE}.

\subsection{Waveforms}

The decrease in oscillation amplitude as the bursts in some sources
progress (Strohmayer et al.\ 1997b) may suggest an 
initially localized emission spot that
expands via relatively slow ($\sim 10^6$cm~s$^{-1}$) thermonuclear propagation.
If so, we would expect that the waveforms from burst oscillations would
reflect a variety of spot sizes.  We therefore consider spots
that range from pointlike to those with an angular radius of 180$^{\circ}$.
Also, physical conditions existing in the region of emitting spots may alter 
photon emission as in the case of some radio pulsars. Accordingly, we consider
the effects of including cosine and sine beaming functions in the calculations
of the waveforms.

\begin{figure*}[!h] 
\psfig{file=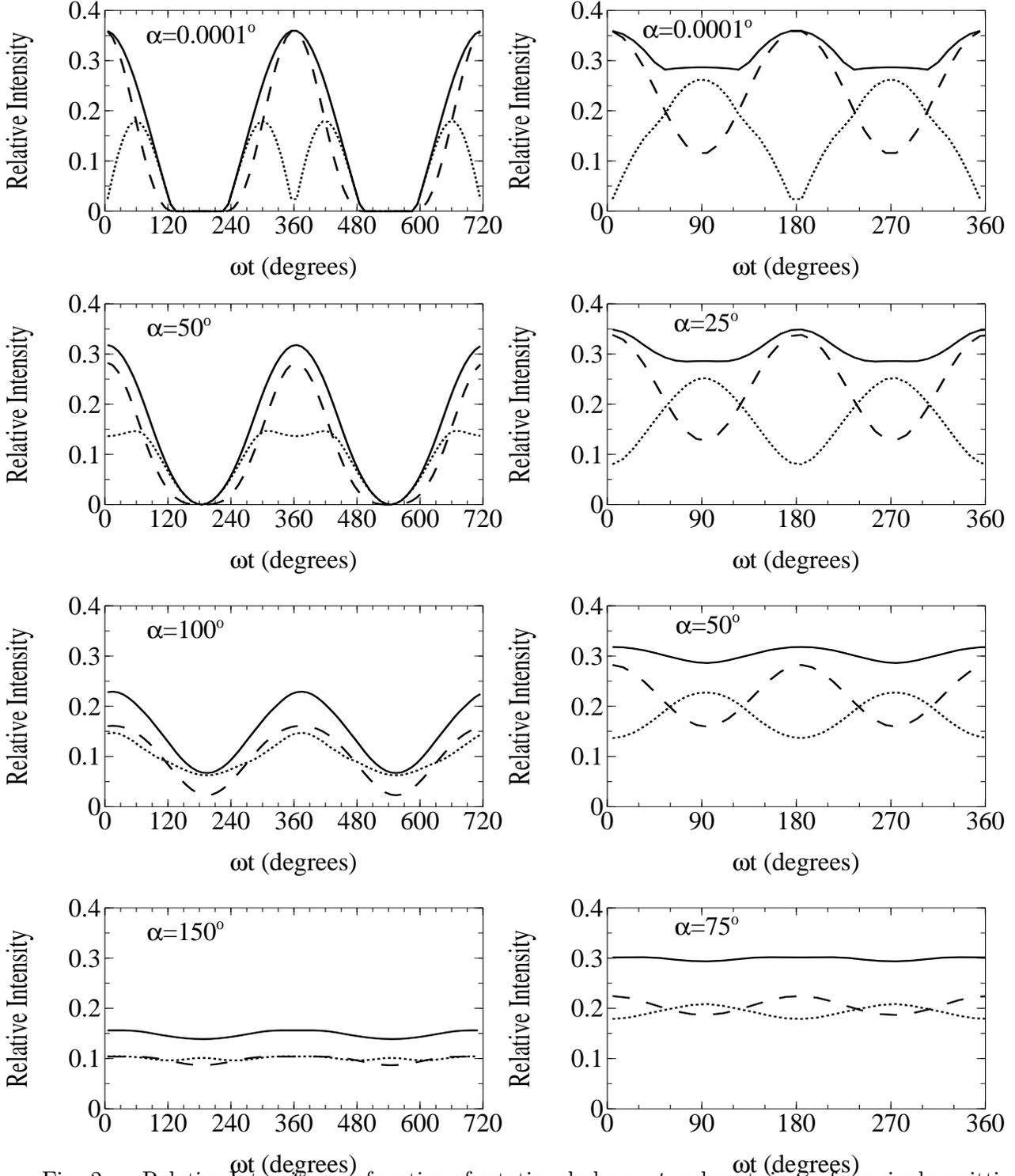,height=8.0truein,width=7.0truein}
\caption[]{\label{waveforms}
Relative Intensity as a function of rotational phase $\omega t$ and spot size 
$\alpha$, for a  single emitting spot (left column) and two emitting
spots (right column)
with $R/M=5.0$ and $\beta=\gamma=90^{\circ}$. Solid line is isotropic 
emission, dashed line is cosine beaming and dotted line is sine beaming. As the
spot sizes increase the amplitude of the intensity oscillations decrease.
Cosine beaming narrows the width of the waveform peaks and sine beaming produces
four peaked waveforms for small single emitting spots.}
\end{figure*}

Figure~\ref{waveforms} 
shows the waveforms from a single emitting spot (left-hand column) and
two emitting spots (right-hand column) for various spot sizes. As expected, the
amplitude of the intensity oscillations decreases as the spot size increases.
Furthermore, in the case of a single emitting spot there is a critical spot size
($\alpha \sim 50^{\circ}$ for the case of $R/M=5.0$) at which the spot is never 
completely out of view and hence the intensity remains greater than zero for 
the entire rotational phase. As the waveforms illustrate, the cosine beaming 
function, which enhances emission along the magnetic field axis, tends to narrow 
the width of the
waveform peaks. The sine beaming function enhances emission near the tangential
plane and will produce a four peaked waveform for the case of a small single
emitting spot (see Pechenick  et al. 1983).  

\subsection{Effects of Spot Size and Light Deflection}

\begin{figure*}[!h] 
\psfig{file=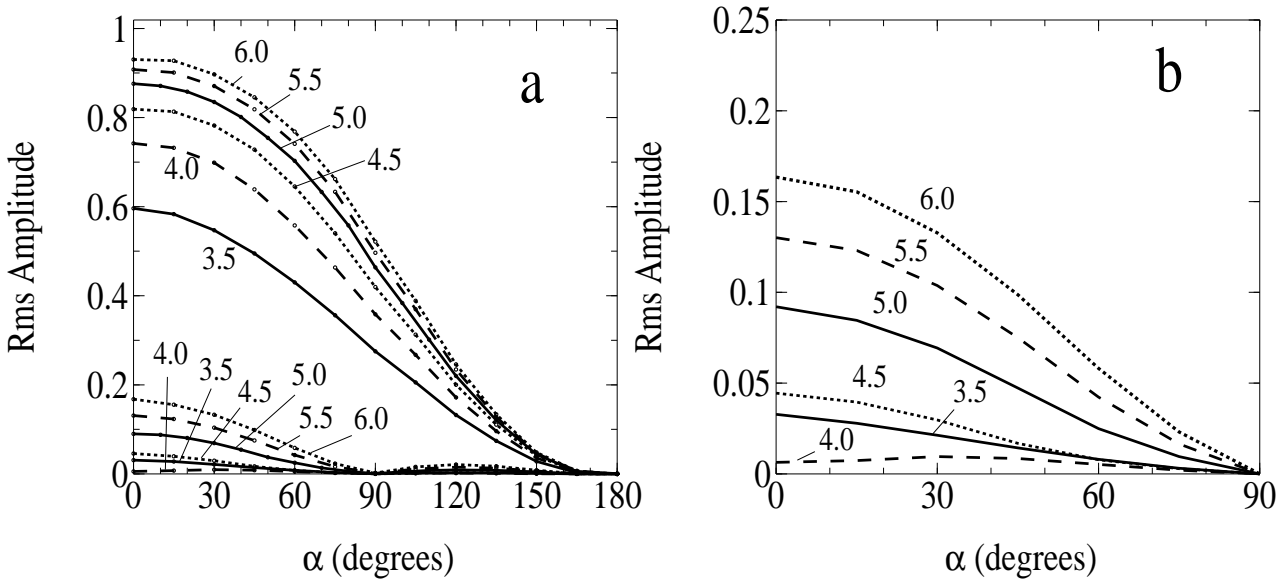,height=4.0truein,width=7.0truein}
\caption[]{\label{ampvssize}($a$)
Rms amplitude as a function of spot size $\alpha$ and stellar compactness 
at the first harmonic (upper curves)  and the
second harmonic (lower curves) from a single emitting spot.  Numbers denote
the value of $R/M$ used for each curve (where we use geometrized units
in which $G=c\equiv 1$), and in each case we assume
$\beta=\gamma=90^\circ$, i.e., that both the observer and the center of
the spot are in the rotational equator.
($b$) Rms amplitude as a function of spot size and stellar compactness at the
second harmonic from two antipodal emitting spots.
Note that the vertical scale is different
than in panel~($a$).  As before, numbers denote $R/M$ and
we assume $\beta=\gamma=90^\circ$.  These figures show that the amplitude
remains relatively unchanged as the spot size increases until 
$\alpha\sim 40^\circ$, at which point the amplitude drops sharply.
It is therefore expected that, if thermonuclear propagation proceeds
at a constant and slow speed, the amplitude versus time curve should
initially exhibit negative curvature.  These figures also show that the
bolometric amplitude at the second harmonic is much more strongly affected
by the stellar compactness than is the amplitude at the first harmonic.} 
\end{figure*}

We are also interested in the effect of the compactness of
the star on the observed amplitudes. Figure~\ref{ampvssize}a shows the 
fractional rms amplitudes at the first two harmonics as a function of spot size 
and stellar compactness for one emitting spot centered at 
$\beta = \gamma = 90^{\circ}$ (i.e., for an observer and spot both in 
the rotational equator). The curves for the first harmonic illustrates  the
general shape of most of the first harmonic curves. Initially, the amplitude
depends only weakly on spot size. However, once the spot grows to $\sim
40^{\circ}$ there is a steep decline in the oscillation amplitude which
flattens out only near the tail of the expansion. Figure~\ref{ampvssize}b 
shows the 
fractional rms amplitude at the second harmonic under the same assumptions but
for two identical, antipodal emitting spots. The range in spot size here is 
$0^{\circ}-90^{\circ}$ since two antipodal spots of $90^{\circ}$ radii cover 
the entire stellar surface. Note that in this situation,  there is no first
harmonic. 

These curves illustrate two interesting features of the two spot
configuration. First, the strength of the strongest oscillation amplitude in
the two spot case is $\sim 90\%$ weaker than the strength of the strongest
oscillation amplitude in the one spot case considered above. Furthermore, the
curve of the second harmonic does not exhibit the same sharp falloff seen in
the first harmonic curve. Thus, the detection of a particularly large
fractional rms  amplitude with a steep amplitude decline  can verify that what
is being observed  is a first harmonic (i.e., power generated at the stellar
spin frequency) rather than any higher harmonics (see Miller \& Lamb 1998). 
The second interesting feature is that the curve of the second harmonic in 
Figure~\ref{ampvssize}b
is nearly identical in both magnitude and shape to the first $90^{\circ}$
of the curve of the second harmonic for the case of one spot shown in 
Figure~\ref{ampvssize}a.
Thus, for this geometry, the introduction of a second emitting spot
antipodal to the first tends to destroy  the first harmonic while leaving the
second harmonic unaffected. This result obtains whenever:
(1) the physical assumptions (e.g.,
compactness, rotational velocity, flux distribution function) made for both the
one and two spot  configurations are the same, and (2) the viewing geometry
for both configurations is $\beta =\gamma = 90^{\circ}$. 

In this figure we also
display the effect gravity has on the oscillation amplitudes.  From equation
(3) we know that more compact stars have a larger $\phi_{max}$, and hence a
larger fraction of their surface is visible to observers. 
As a result, oscillation
amplitudes for more compact neutron stars are smaller. An exception occurs at
the second harmonic  of very compact stars ($R/M < 4.0$), in which case
gravitational light deflection focuses the emitted radiation enough to raise the
oscillation amplitude (see Pechenick  et al. 1983 and Miller \& Lamb 1998).
Note that the stellar compactness affects the amplitude at the second
harmonic far more than the amplitude at the first harmonic.

\subsection{Effects of Viewing Angle and Magnetic Inclination}

\begin{figure*}[!h] 
\psfig{file=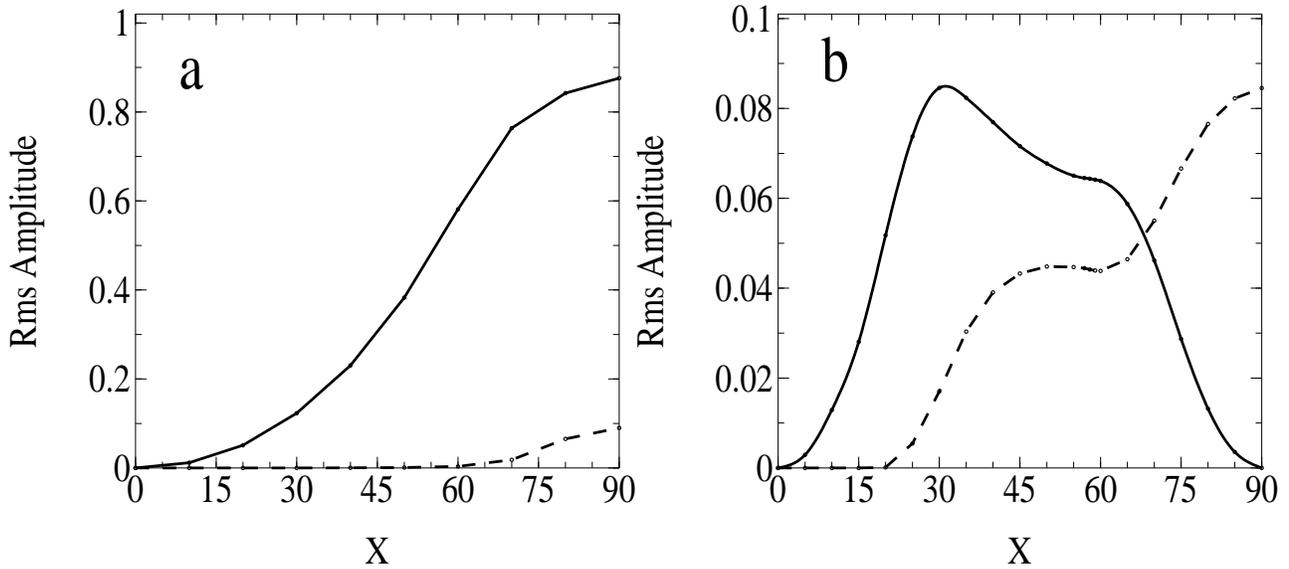,height=4.0truein,width=7.0truein}
\caption[]{\label{ampvsang}
Rms amplitude as a function of $\beta=\gamma=X$ (i.e., for an
observer at the same angular distance from the rotation as the hot
spot is).  ($a$)~Single
emitting spot with $R/M = 5.0$ and $\alpha = 15^{\circ}$.  The solid
line plots the amplitude of the first harmonic, and the dashed line plots
the amplitude of the second harmonic.  This panel demonstrates that if there
is only one spot, the presence of a second harmonic means that the line
of sight of the observer cannot be close to face-on.
($b$)~Two antipodal emitting spots, where again $R/M = 5.0$ and 
$\alpha = 15^{\circ}$, and the amplitudes of both the first harmonic
(solid line) and second harmonic (dashed line) are plotted.  This panel
shows that as the line of sight and location of the hot spot move from
the rotational axis to the rotational equator, the power in the first
harmonic is gradually transferred to the second harmonic.}
\end{figure*}

Figure~\ref{ampvsang}a shows the
oscillation amplitude as a function of $\beta = \gamma = x $ (i.e., for
the observer and the center of the spot at the same rotational latitude)
for a single
emitting spot with $\alpha = 15^{\circ}$  and $R/M = 5.0$. As $x$ increases,
the width of the peaks in the light curve decrease (see Pechenick et al.\ 1983)
and hence the oscillation amplitudes increase.  The interesting feature here is
that the second harmonic has a significant amplitude only for $x >
60^{\circ}$. Since 50\% of the time $x$ will be between $60^{\circ}$ and
$90^{\circ}$ (assuming randomly distributed observers), only half of all
observers will detect a second harmonic during a typical burst involving one
spot. In Figure~\ref{ampvsang}b we make the same assumptions  as in 
Figure~\ref{ampvsang}a but for
two emitting spots rather than one. If we had assumed flat space-time and an
infinitesimal spot size then the second emitting spot would  become visible
only for $2x = 180^{\circ} - \phi_{max} = 180^{\circ} - 90^{\circ}  =
90^{\circ}$. Therefore, for $x < 45^{\circ}$ only one spot would be observable.
For $R/M = 5.0$,
$\phi_{max} = 128^{\circ}$, and therefore a second, infinitesimal, spot would
begin to be visible at $x=26^{\circ}$. Since in Figure~\ref{ampvsang}, 
the calculation
was done with  $\alpha = 15^{\circ}$, the spot begins to become visible at $x =
26^{\circ} - (15/2)^{\circ} \approx 20^{\circ}$, explaining the appearance of
the second harmonic at this $x$ value.  Note that in the two spot case the
first harmonic generates significant power for a wide range of $x$. This
occurs because for $x \ne 90^{\circ}$ one spot is more directly aligned with
the observer's line of sight, and as a result the intensity maxima of the two
spots are unequal. In general, whenever an asymmetry exists between the two
emitting spots such that the intensity maxima of the two spots are unequal, 
power is generated at the first harmonic.

\subsection{Effects of Anisotropies from Doppler Shifts and Beaming}

\begin{figure*}[!h] 
\hbox{\hskip 1.2truein \vbox{\vskip -0.7truein
\psfig{file=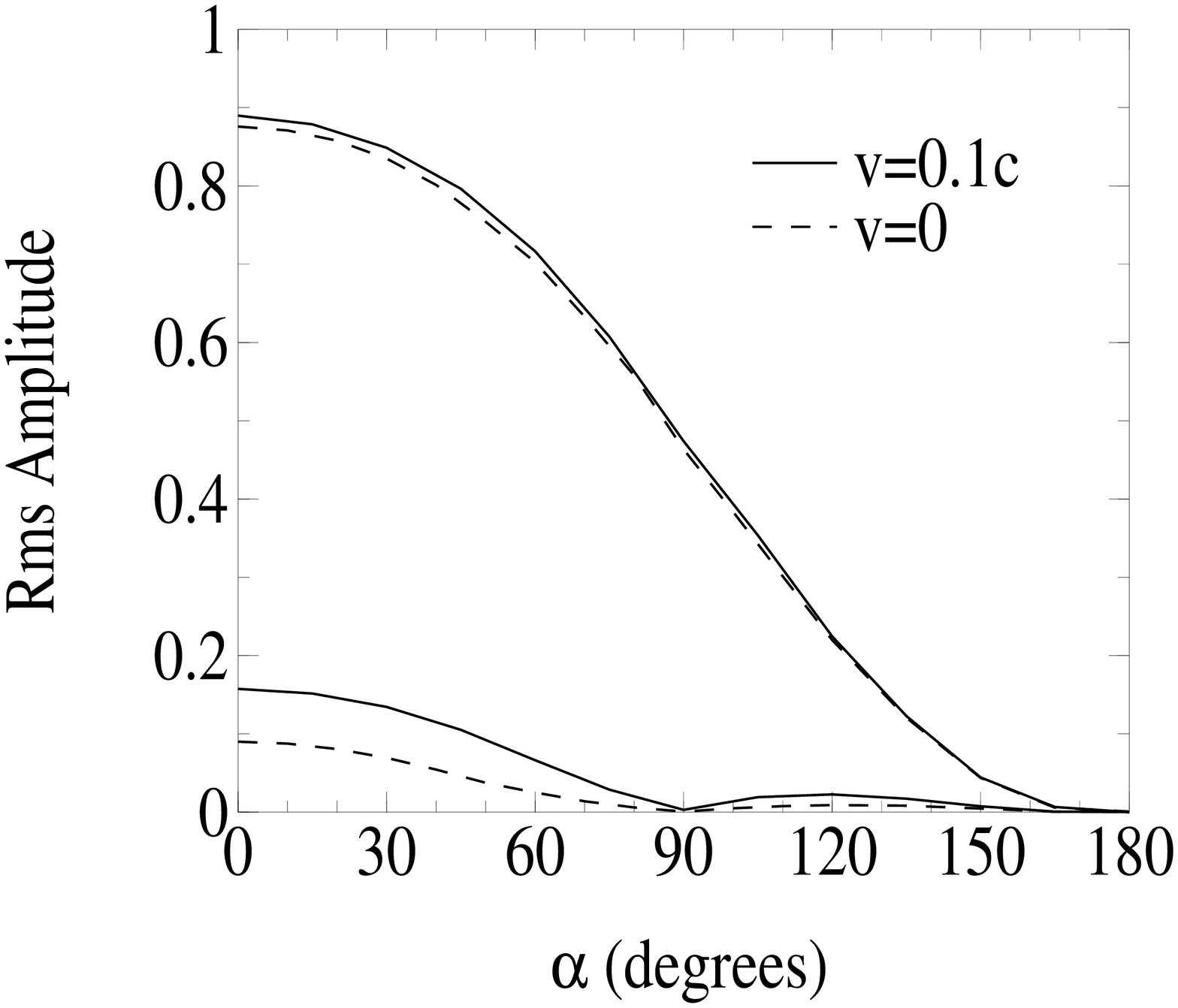,height=4.0truein,width=4.0truein}}} 
\caption[]{\label{ampvsvel}
Effect of surface rotational velocities on different harmonics.
Here we plot rms amplitude vs. spot angular radius 
$\alpha$  at the first harmonic (upper curves) and the  second
harmonic (lower curves) from a single emitting spot with a surface 
velocity measured at infinity of $v=0.1c$ 
(\textit{solid lines}) and $v=0.0c$ (\textit{dashed lines}), $R/M = 5.0$, and
$\beta = \gamma = 90^{\circ}$ (i.e., observer and hot spot both in the
rotational equator).  The increase in amplitude due to nonzero rotational
velocities is much greater for overtones than it is for the fundamental
of the stellar spin frequency.} 
\end{figure*}

In Figure~\ref{ampvsvel} we include the effects of
Doppler shifts and aberrations on the  oscillation amplitudes. We assume
a surface rotation velocity of $v= 0.1$c, which corresponds to
a neutron star with radius $R = 10$~km and spin frequency $\nu \approx 400$~Hz.
As can be seen, the amplitude of the second harmonic is increased significantly
more than the amplitude of the first harmonic as a result of rotation.  The
tendency to generate more power at the higher harmonics than at the spin
frequency is a general property of the rotation (see Miller \& Lamb 1998 for a
discussion of this effect).

\begin{figure*}[!h] 
\psfig{file=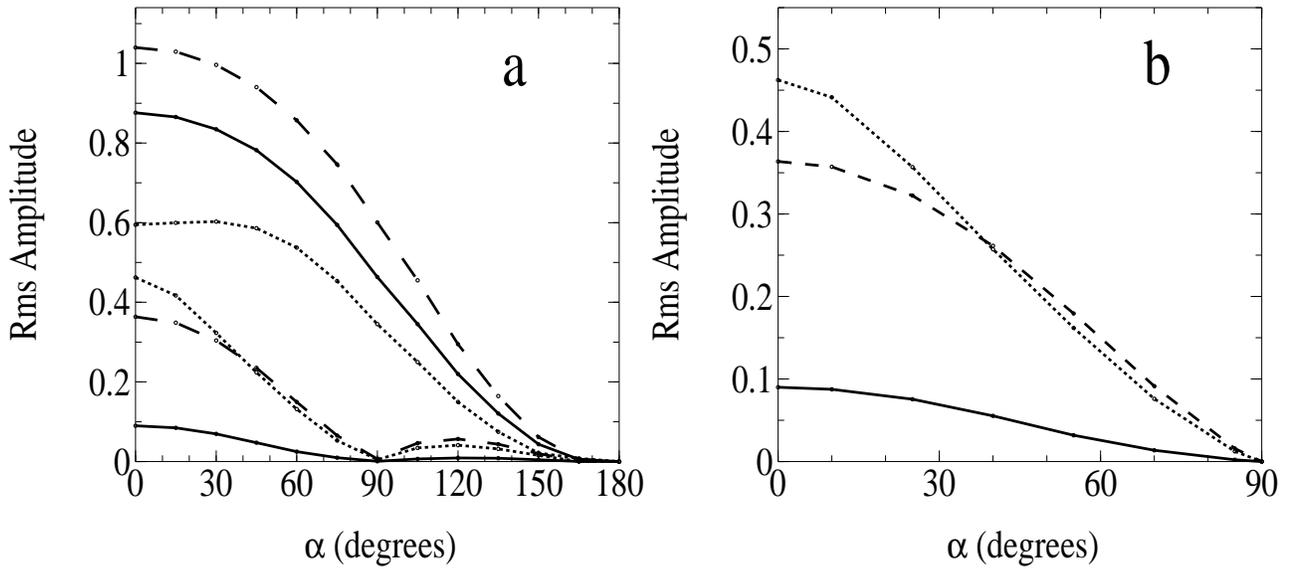,height=4.0truein,width=7.0truein}
\caption[]{\label{ampvsbeam}
Effect of different beaming functions on the first and second
harmonics.  ($a$)~Rms amplitude
vs. $\alpha$ at the first harmonic (upper three curves) and the
second harmonic (lower three curves) from a single emitting spot  with $R/M =
5.0$, $\beta = \gamma = 90^{\circ}$. Dashed line is cosine beaming, solid line
is isotropic emission, and dotted line is sine beaming. ($b$)~Rms amplitude vs.
$\alpha$ at the second harmonic from two emitting spots with $R/M = 5.0$,
$\beta = \gamma = 90^{\circ}$. Again, dashed line is cosine beaming,  solid
line is isotropic emission, and dotted line is sine beaming.  Both of the
anisotropic emission patterns enhance tremendously the amplitude at the
second harmonic, by a factor of at least 4.} 
\end{figure*}

Physical conditions in the region of emitting spots might cause anisotropic
emission of radiation. The results of including a cosine beaming function and a
sine beaming function for the case of one spot are shown in 
Figure~\ref{ampvsbeam}a and
for two antipodal spots in Figure~\ref{ampvsbeam}b. As is apparent from 
Figure~\ref{waveforms}, the 
enhanced emission along the
magnetic axis for the cosine beaming tends to narrow peaks in the light curves
(see Pechenick et al.\ 1983 for a discussion of the light curves for beamed
emission) and hence raise the oscillation amplitudes. For the sine beaming  the
peaks in the light curve are broadened, tending to lower the amplitude at the
first harmonic. Both beaming functions do, however, generate substantial
additional power at the higher harmonics.

\section{APPLICATION TO X-RAY BURST SOURCES}

\subsection{Relative Amplitudes of Harmonics in 4U~1636--536}

\begin{figure*}[!h] 
\begin{minipage}[t]{3.1truein}
\mbox{}\\
\psfig{file=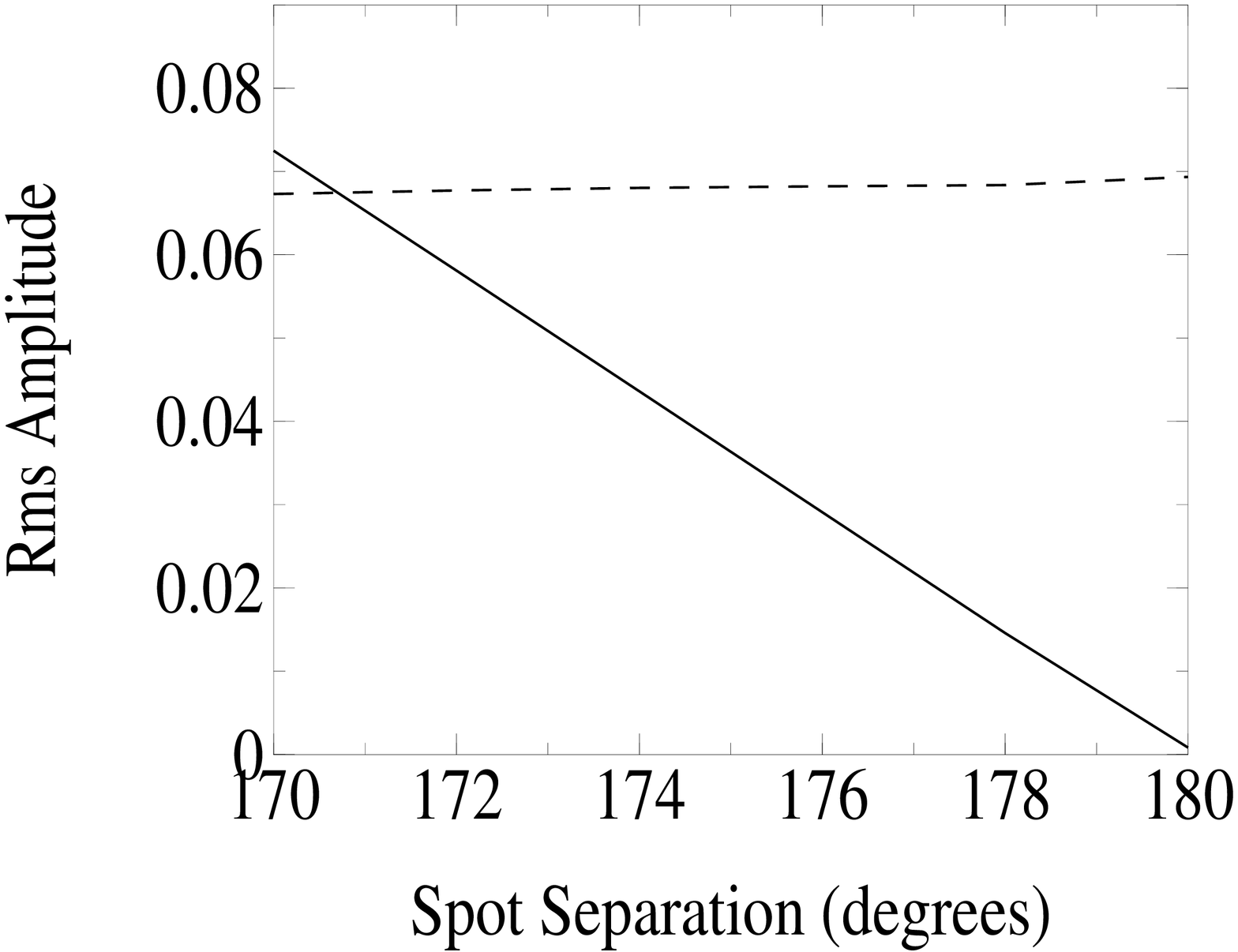,height=2.9truein,width=2.9truein}
\end{minipage}
\begin{minipage}[t]{3.1truein}
\mbox{}\\
\psfig{file=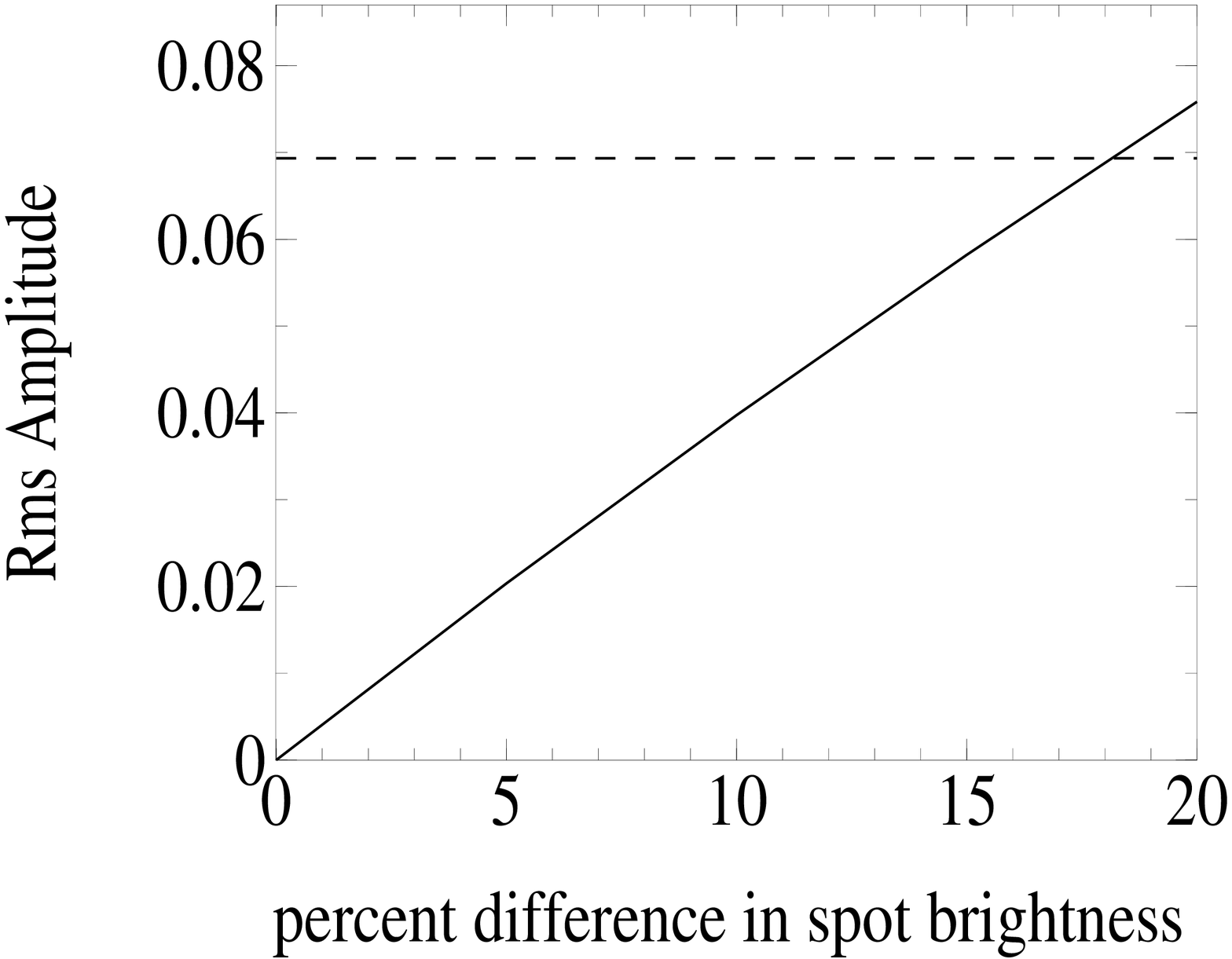,height=2.9truein,width=2.9truein}
\end{minipage}
\caption[]{\label{asymmetric}
Effect of nonantipodal separation of spots and differences
in spot brightness on the relative amplitudes at the first and second
harmonics. (left panel) Rms
amplitude vs. spot separation at the first harmonic (\textit{solid line}) and
the second harmonic (\textit{dashed line}) for two emitting spots with $R/M =
5.0$,  $\alpha = 30^{\circ}$, $\beta = \gamma = 90^{\circ}$.
(right panel) Rms
amplitude vs. percent difference in spot brightness at the first harmonic
(\textit{solid line}) and the second harmonic (\textit{dashed line}) for two
antipodal emitting spots with $R/M = 5.0$, $\alpha = 30^{\circ}$, $\beta =
\gamma = 90^{\circ}$.  The analysis of RXTE data from 4U~1636--536 by
Miller (1999) indicates that the average ratio of amplitudes of the second
to first harmonic is 2.3.  This figure therefore places strong constraints
on the spot separation and possible differences in spot brightness: the
spot separation in 4U~1636--536 must be within 4$^\circ$ of antipodal,
and the spot brightnesses must be within 7\% of each other, or else the
amplitude at the fundamental of the spin frequency would be larger than
observed relative to the overtone.} 
\end{figure*}

Recent work by Miller (1999) gives
evidence for the presence of power at the stellar spin frequency for a source
(4U~1636--536) consisting of two emitting spots.  Earlier we saw that one
possible mechanism for generating significant power at the stellar  spin
frequency for the case of two emitting spots is to vary the viewing geometry.
Another possible mechanism is to have the spots be non-antipodal. This can
occur, for instance, if the star's dipolar magnetic field has its axis
slightly displaced from its center.  In the left panel of
Figure~\ref{asymmetric} we show the oscillation
amplitude as a function of spot separation for the case of two emitting spots
with $\alpha = 30^{\circ}$, $\beta = \gamma = 90^{\circ}$,  and $R/M = 5.0$.
The spots are perfectly antipodal at a spot separation of $180^{\circ}$.  As
the figure shows, the oscillation amplitude at the second harmonic is
relatively constant while the oscillation amplitude at the first harmonic is a
linear function of spot separation. At a spot separation of 
$\sim170^{\circ}$ the
fractional rms amplitudes of the first and second harmonic are equal. Another
way to produce power at the spin frequency is to have
differences in brightness between the two spots. Such an asymmetry can occur,
for example, if the strength of the magnetic field at the location of the two
spots is different, thereby pooling different amounts of nuclear fuel  onto the
hot spot regions. In the right panel of 
Figure~\ref{asymmetric} we show the oscillation amplitude as a function
of the percent difference between the brightness of the two spots. As in the
case of the non-antipodal spots, the amplitude at the second harmonic is
essentially constant while the amplitude at the first harmonic increases
linearly with increasing percent difference in spot brightness. 

These figures reinforce the conclusion, also evident from 
Figure~\ref{ampvssize}, that only with two spots can the oscillation
at the first overtone be stronger than the oscillation at the
fundamental.  Therefore, within the general theoretical model explored
in this paper, 4U~1636--536 has two nearly antipodal hot spots.

\subsection{Phase Lags in SAX~J1808--3658}

{\it Doppler model of phase lags.}---The hard X-ray spectrum of low-mass 
X-ray binaries is well-fit by a
Comptonization model, in which the central neutron star is surrounded
by a hot corona of electrons and the photons
injected into this corona are relatively soft.
It was therefore expected that the observed hard photons, having 
scattered more often than the soft photons and thus having a longer
path length before escape, would lag the soft photons.  Instead, in
several sources a hard {\it lead} was discovered.  One explanation for
this lead was suggested by Ford (1999).  He proposed that
Doppler shifting of photons emitted from rotating hot spots, as in
thermonuclear burst oscillations, would tend to produce a hard lead
because the approaching edge of the spot would precede the trailing
edge.  He compared a simplified calculation of this effect with burst
data for Aql~X-1 and showed that an adequate fit could be obtained (Ford 1999).

The millisecond X-ray pulsar SAX~J1808--3658 provides a stronger test
of this hypothesis.  This source has strong oscillations ($\sim$5\% rms)
at $\sim$401~Hz, which as usual are attributed to rotational modulation
of a hot spot on the surface.  Cui, Morgan, \& Titarchuk (1998) obtained 
precise measurements of the oscillation phase as a function of energy, and 
found that in this source as well there is a hard lead.  

\begin{figure*}[t]
\begin{minipage}[t]{3.1truein}
\mbox{}\\
\psfig{file=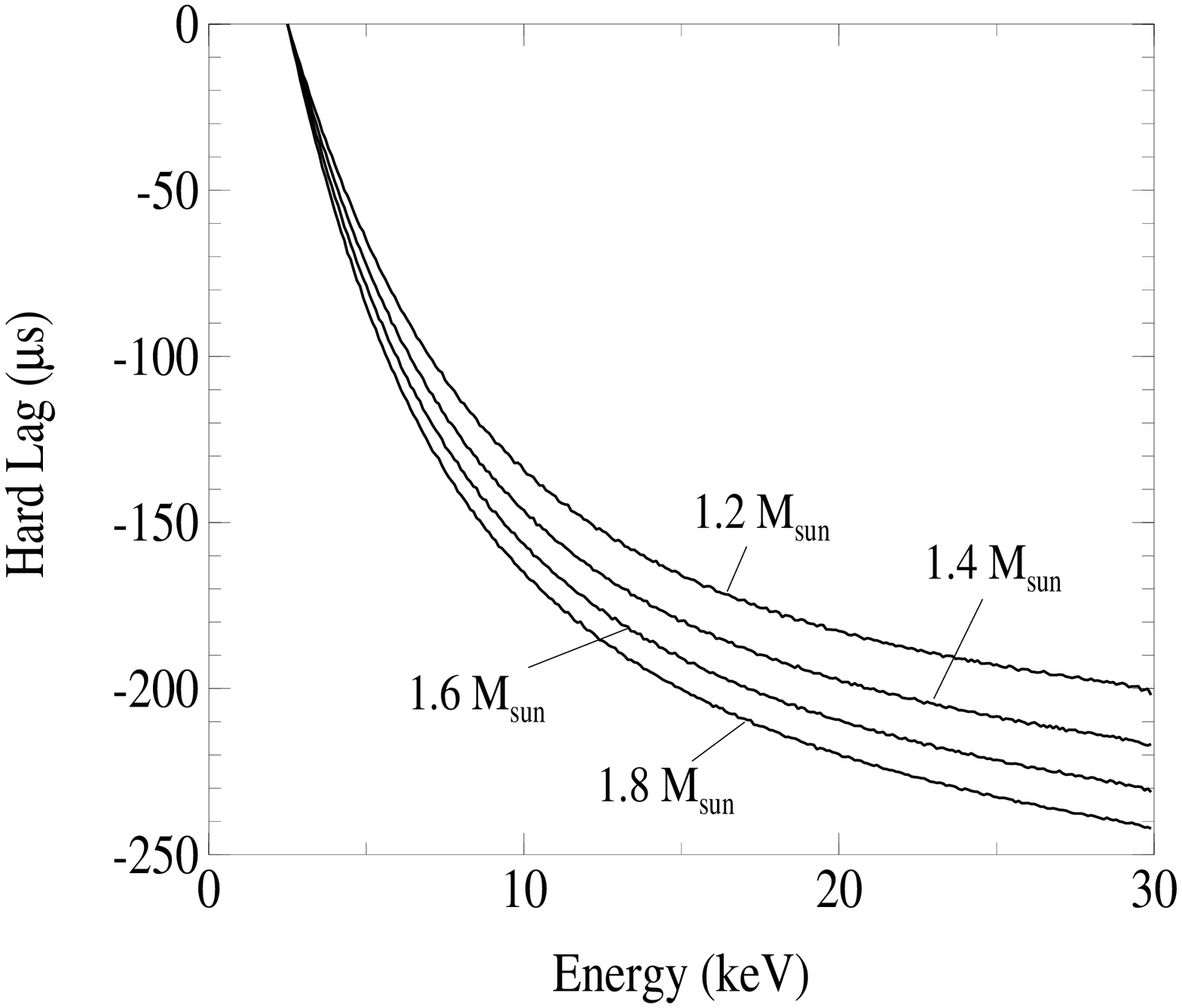,height=2.9truein,width=2.9truein}
\end{minipage}
\begin{minipage}[t]{3.1truein}
\mbox{}\\
\psfig{file=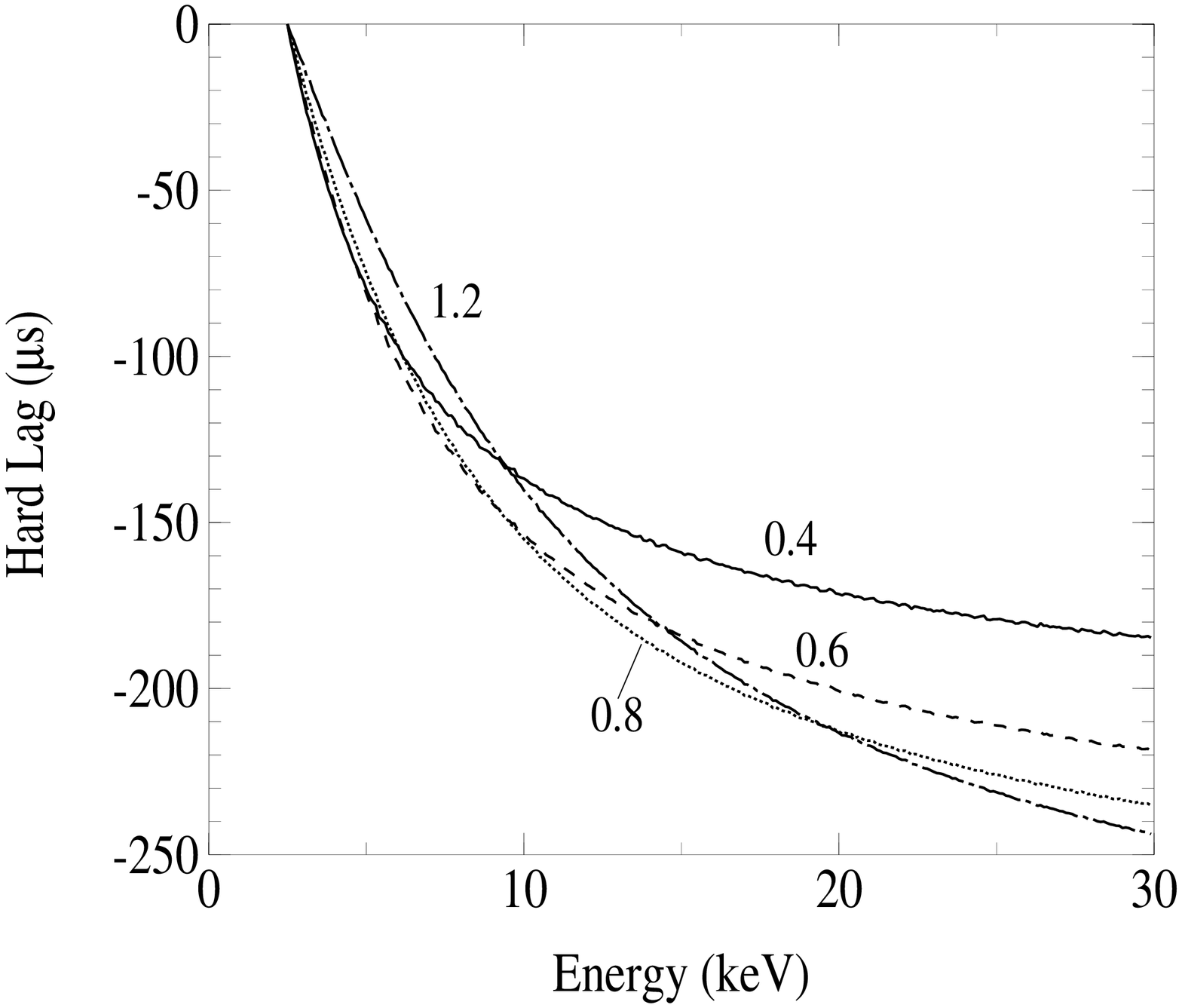,height=2.9truein,width=2.9truein}
\end{minipage}
\caption[]{\label{lagfigs}
Time lags versus photon energy, as a function of stellar
gravitational mass (left panel) and of surface temperature (right panel).
The lags are relative to the photons at 2.5~keV, and their negative
value indicates a hard lead.
In both cases we assume a stellar spin frequency of 401~Hz to correspond
to the spin frequency of SAX~J1808--3658.  We also assume a stellar
compactness of $R/M=5.1$, and assume that the surface emission has the
pattern appropriate for a gray atmosphere.  In the left panel the curves are
labeled by the gravitational mass, and we assume a surface effective
temperature of $kT=0.7$~keV as measured at infinity.  In the right panel the
curves are labeled by the surface effective temperature (as measured
at infinity) in units of keV, and we assume $M=1.6\,M_\odot$.}
\end{figure*}

Figure~\ref{lagfigs} shows sample calculations of the time lag 
as a function of 
energy.  In the left panel we focus on the dependence of the lag on mass,
and in the right panel we concentrate on the effect of changing the surface
temperature.  In both cases the surface emission pattern is the pattern
for a gray atmosphere, and we assume $R/M=5.1$ and a stellar spin
frequency of 401~Hz, which is the spin frequency of SAX~J1808--3658.
In panel~(a) we assume a surface effective temperature of $kT=0.7$~keV
as measured at infinity.  In panel~(b) we assume a stellar gravitational
mass of 1.6~$M_\odot$, which gives a surface equatorial rotation velocity
of 0.1~c as measured at infinity.  From this figure it is clear that the
effect of increasing the mass is to increase the phase lead, whereas the
effect of increasing the temperature is to increase the energy at which
the curve starts to flatten.

\begin{figure*}[t]
\psfig{file=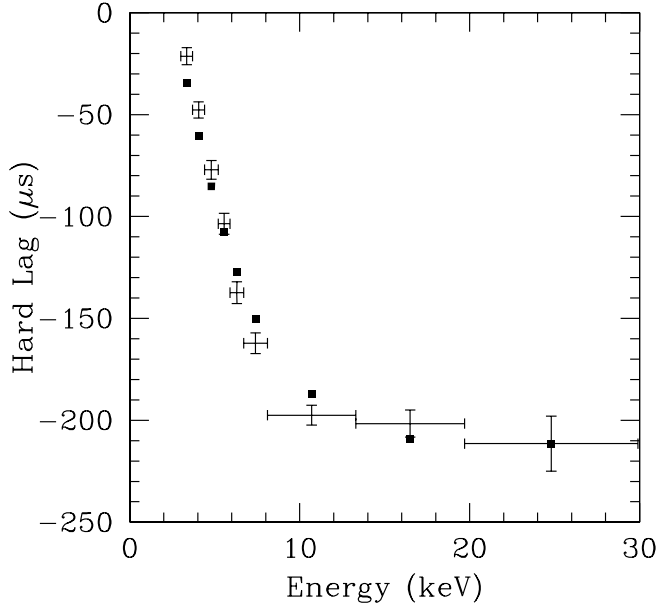,height=3.5truein,width=3.5truein}
\caption[]{\label{1808data}
Comparison of model time lags with the lags reported by Cui et al.\ (1998)
for SAX~J1808--3658.  The vertical axis is the time lag in microseconds
relative to the average in the 2--3~keV band, and the horizontal axis
is the observed photon energy in keV.  The crosses are the data: the 
horizontal bars indicate the extent of each energy bin, whereas the 
vertical bars indicate the uncertainty in the time lag.  The model time
lags are shown with the filled boxes, and are computed via the procedure
described in the text.  In this fit, the neutron star gravitational mass
is $M=2.2\,M_\odot$, $R=10$~km, the surface temperature measured at 
infinity is $kT=1.1$~keV, and we assumed an isothermal atmosphere.  The
total $\chi^2$ was 38.6 for six degrees of freedom.  The reasonable 
quantitative fit to the data add support to the Doppler shift hypothesis
for the origin of the hard leads.  The data for SAX~J1808--3658 were kindly 
provided by Wei Cui.}
\end{figure*}

{\it Comparison with data.}---Comparing these models with the data for
SAX~J1808--3658 introduces additional complications.  In order to improve
statistics, Cui et al.\ (1998) averaged the phase lags over the period
from 11 April 1998 to 29 April 1998.  The calculation
of the phase leads by Cui et al.\ (1998) also involves averaging the phase
over energy bins several keV in width.  Examination of Figure~\ref{1808data}
shows
that the phase changes rapidly over such an energy range, implying
that the measured phase lead depends sensitively on the input spectrum.
The effective area of RXTE also decreases rapidly below 4~keV, which
strongly affects the observed average phase in the 2--3~keV reference
bin.  Finally, given that the observed spectrum is not a blackbody, but
is instead approximately a power law of index 1.86 from $\sim$3~keV
to $\sim$30~keV (Heindl \& Smith 1998), Compton reprocessing has taken
place and the observed phase lags are the result of a convolution between
the unscattered phase lags and the Compton redistribution function.

Figure~\ref{1808data} plots the data along with a 
simplified model of the phase lags
taking some of these complications into account.  We ignore the changing
effective area of RXTE and assume a constant response with energy.  Based
on the power-law nature of the spectrum, we approximate the process of
Comptonization by assuming that the energy of the injected photons is
much less than the observed photon energies or the temperature of the
electrons.  We also assume an isothermal atmosphere, in contrast to the
gray atmosphere we used for Figure~\ref{lagfigs}, which gives too low
a hard lead.  The best fit has $kT=1.1$~keV
as observed at infinity, $R=10$~km, and $M=2.2\,M_\odot$.  The total $\chi^2$
of the fit is 38.6 for 6 degrees of freedom.  The dominant contribution to
this $\chi^2$ comes from the underprediction of the hard lead at low
energies.  This is as expected, because we assumed an instrumental effective
area that is constant with energy, whereas in reality the effective area
rises rapidly with increasing photon energy at low energies.  This changing
effective area gives greater weight to the larger leads at higher energies,
which is in better agreement with the data.  Therefore,
given the simplifications of the model, our fit to the data
is encouragingly good and supports the Doppler interpretation
of the observed hard lead.

\section{DISCUSSION}

{\it Relative amplitudes of harmonics.}---We have presented 
calculations of the waveforms and amplitudes at
different harmonics of the spin frequency for one or two hot spots
and many realistic combinations of stellar compactness, spot size
and emission pattern, observation angle, and magnetic inclination.
These calculations show that typically either the fundamental or the
first overtone has an amplitude much larger than the amplitude of
any other harmonic.  This corresponds well to the observations of
the six sources with burst brightness oscillations, in which there
is a strong oscillation at only one frequency.  We also find that
if the first overtone is the dominant harmonic, there must be two
similar and nearly antipodal bright spots, because a single spot
always produces a much stronger oscillation at the fundamental than
at any overtone.  In contrast, if the fundamental is much stronger
than the overtone, this is consistent with but does not require a
single spot: if there are two bright spots that are sufficiently
dissimilar or far away from antipodal, or if our line of sight is such
that one of the spots is hidden, then the oscillation at the
fundamental will dominate.  This implies that the three sources with
detectable oscillations near $\sim$300~Hz (4U~1728--34 [Strohmayer et
al.\ 1996],
4U~1702--43 [Markwardt, Strohmayer, \& Swank 1999], and 
4U~1636--536 [Miller 1999; this source
has a strong oscillation at $\sim$580~Hz but a detectable oscillation
at $\sim$290~Hz]) have spin frequencies of $\sim$300~Hz, whereas the
three sources with detectable oscillations only at $\sim$500--600~Hz
(Aql~X-1 [Zhang et al.\ 1998], MXB~1743--29 [Strohmayer et al.\ 1997a], 
and KS~1731--260
[Smith et al.\ 1997]) could have spin frequencies at either this frequency or
half of it.  Therefore, all six burst oscillation
sources are consistent with having spin frequencies $\sim$300~Hz.

\begin{figure*}[t]
\psfig{file=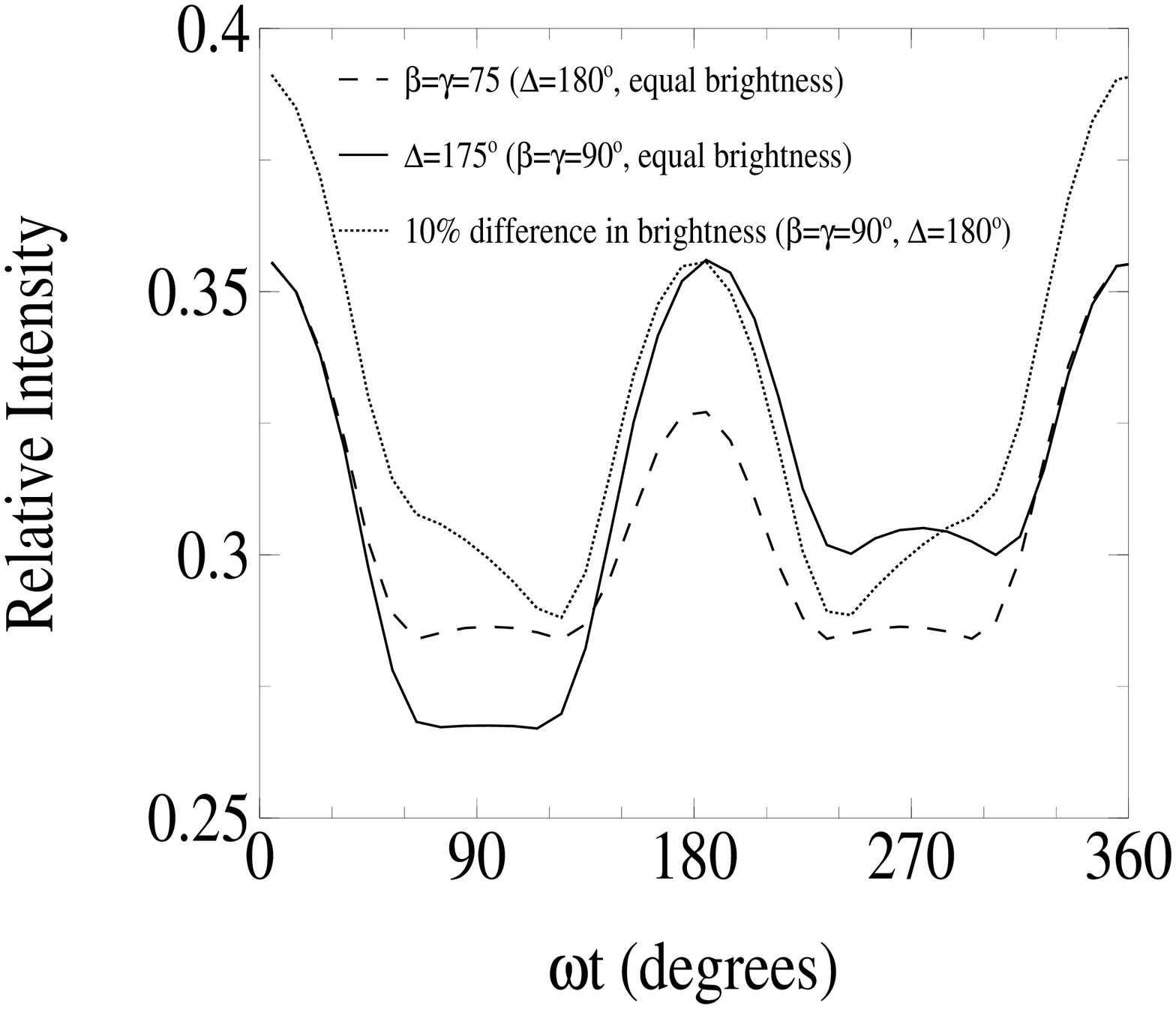,height=3.5truein,width=4.5truein}
\caption[]{\label{fixedratio}
Burst oscillation waveforms with a fixed ratio between the amplitude at
the first overtone and the amplitude at the fundamental of 2.3, which
was the ratio found by Miller (1999) for 4U~1636--536.  Solid line:
identical spots 175$^\circ$ apart.  Dotted line: antipodal spots differing
in brightness by 10\%.  Dashed line: identical, antipodal spots 75$^\circ$
from the rotation axis as seen by a distant observer 75$^\circ$ from the
rotation axis.  For all three curves, $R/M=5$  and the spot angular
radius is 15$^\circ$.  This figure shows that amplitude ratios, as are 
computed with a power density spectrum, are not sufficient to distinguish 
between these various scenarios.  Waveforms are to be preferred.}
\end{figure*}

{\it Information content of waveforms.}---Our results also show clearly
that power density spectra, which contain information only on the
relative amplitudes of different harmonics, are much less informative than
the waveforms themselves.  Figure~\ref{fixedratio} shows three 
different waveforms
that all have an amplitude at the first overtone that is 2.3 times the
amplitude at the fundamental, which is the ratio found by
Miller (1999) for 4U~1636--536.  In all three cases there are two bright spots.
The solid line shows the waveform for two identical pointlike spots that
are 175$^\circ$ apart, the dotted line shows the waveform for two antipodal
spots with brightnesses differing by 10\%, and the dashed line shows the
waveform for two identical and antipodal spots that are 75$^\circ$ from
the rotational pole and observed from a line of sight that is also
75$^\circ$ from the rotational pole.  Although the amplitude ratio is
the same in each case, the waveforms are quite different from each other,
and the physical implications are also different.
This underscores the importance of calculating waveforms and not just
power density spectra, both observationally and theoretically.

{\it Searches for weak higher harmonics.}---The amplitudes and phases of
higher harmonics potentially contain important clues about the propagation
of nuclear burning and about the compactness of the star, but as yet there
are no sources in which a higher harmonic of a strong oscillation has been
observed.  Our plots of amplitude versus spot size suggest 
that it is best to look for weaker
higher harmonics when the dominant oscillation is strong.  The reason is
that, in general, the 
ratio of the second to first harmonic drops with increasing spot size, and
therefore with decreasing amplitude at the fundamental.  Hence, a search
only of data showing strong oscillations may more sensitively reveal the
presence of higher harmonics.

{\it Shape of amplitude decrease as a function of spot size.}---In our
calculations, as the spot size increases the amplitude decreases slowly
until the angular radius of the spot is $\sim 40^\circ$ but quickly
thereafter.  This apparently conflicts with the observations of 4U~1728--34
reported in Strohmayer et al.\ (1997), in which the error bars are large
but it appears that the decrease in amplitude is fast from the start and
then slows down.  Further quantification of this result is important, but
if confirmed it could be caused by a number of effects.  For example,
the spot size might never be small: if ignition were nearly simultaneous
over a large area, further spreading would already be in the large-spot
regime, and hence the amplitude would decrease quickly.  If the spreading
velocity were initially high but then decreased, this would have a similar
effect on the amplitudes.  Alternately, if there is a corona with a
non-negligible scattering optical depth around the star and the optical
depth increases as the burst approaches its peak flux, this would also
decrease the amplitude faster than expected when the optical depth is zero.

{\it Phase lags as a probe of surface rotation velocity.}---We find that the
hard lead observed in SAX~J1808--3658 is fit reasonably well by a model
(see Ford 1999) in which rotational Doppler shifts cause higher-energy
X-rays to lead lower-energy X-rays.  This fit lends support to the model,
and suggests that with better fitting and more data (especially from a
future high-area timing mission) it may be possible to use phase lag versus
energy data to help constraint the mass $M$ or the compactness $R/M$ of the
star.

\acknowledgements
We thank Wei Cui for providing time lag data for SAX~J1808-3658.  This
work was supported in part by NASA ATP grant number NRA-98-03-ATP-028,
NASA contract NASW-4690, and DOE contract DOE B341495.

\newpage

\end{document}